\documentclass{elsarticle}

\journal{New Astronomy}
\begin{document}
\begin{frontmatter}
\title{Cosmic rays: extragalactic and Galactic}

\author[Fian,MIPT]{Ya. N. Istomin},
\ead{istomin@lpi.ru}

\address[Fian]{P.~N.~Lebedev Physical Institute,
Leninsky Prospect 53, Moscow, 119991 Russia}
\address[MIPT]{Moscow Institute Physics and Technology, Institutskii per. 9, Dolgoprudnyi, Moscow region, 141700 Russia}

\begin{abstract}

From the analysis of the flux of high energy particles, $E>3\cdot
10^{18}eV$, it is shown that the distribution of the power density
of extragalactic rays over energy is of the power law, ${\bar
q}(E)\propto E^{-2.7}$, with the same index of $2.7$ that has the
distribution of Galactic cosmic rays before so called 'knee',
$E<3\cdot 10^{15}eV$. However, the average power of extragalactic
sources, which is of ${\cal E}\simeq 10^{43}erg \,s^{-1}$, at
least two orders exceeds the power emitted by the Galaxy in cosmic
rays, assuming that the density of galaxies is estimated as
$N_g\simeq 1 Mpc^{-3}$. Considering that such power can be
provided by relativistic jets from active galactic nuclei with the
power ${\cal E}\simeq 10^{45} - 10^{46} erg \, s^{-1}$, we
estimate the density of extragalactic sources of cosmic rays as
$N_g\simeq 10^{-2}-10^{-3}\, Mpc^{-3}$. Assuming the same nature
of Galactic and extragalactic rays, we conclude that the Galactic
rays were produced by a relativistic jet emitted from the Galactic
center during the period of its activity in the past. The remnants of
a bipolar jet are now observed in the form of bubbles of
relativistic gas above and below the Galactic plane. The break,
observed in the spectrum of Galactic rays ('knee'), is explained
by fast escape of energetic particle, $E>3\cdot 10^{15}eV$, from
the Galaxy because of the dependence of the coefficient of
diffusion of cosmic rays on energy, $D\propto E^{0.7}$. The
obtained index of the density distribution of particles over
energy, $N(E)\propto E^{-2.7-0.7/2}=E^{-3.05}$, for $E>3\cdot
10^{15}eV$ agrees well with the observed one, $N(E)\propto
E^{-3.1}$. Estimated time of termination of the jet in the Galaxy
is $4.2\cdot 10^{4}$ years ago.

\end{abstract}


\end{frontmatter}

\section{Introduction}

The main hypothesis of the origin of cosmic rays in the Galaxy is
the acceleration of charged particles to high energies on the
fronts of shock waves formed by supernova explosions. To
ensure the power of cosmic rays, observed in the Galaxy, which
equals about $10^{41} erg \, s^{-1}$, it is necessary to
transform to accelerated particles about $15\%$ of the kinetic
energy of expanding shock waves. These strong shocks produce the
universal spectrum of particle energy distribution, $N(E)\propto
E^{-2}$. This allows to put together cosmic rays originating from
different supernova, and forms the unique particle spectrum,
extending from energies in a few units of $GeV$ to energies
$\simeq 10^{18}eV$. The observed spectrum $N(E)$ significantly
differs from universal, that is explained by the propagation of
cosmic rays in the Galaxy. It has the character of diffusion in
space because of the scattering of charged particles by
magnetic field inhomogeneities. The diffusion coefficient
increases with particle energy, i.e. the lifetime of the fast
particles in the Galaxy decreases. Therefore, the observed
spectrum over energy differs from the spectrum given by sources,
remaining a power law, $N(E)\propto E^{-\beta}$. But the value
of $\beta$ is not constant, it changes from $\beta_1=2.7$ for
$E<3\cdot 10^{15}eV$ to $\beta_2=3.1$ for $E>3\cdot 10^{15}eV$.
The important circumstance is the fact that the spectrum at high
energies becomes softer, not harder. This suggests that the source
of cosmic rays at energies $E<3\cdot 10^{15}eV$ and at energies
$E>3\cdot 10^{15}eV$ is single. It is possible the superposition
of two independent sources, if it would be vice versa,
$\beta_2<\beta_1$. But it is unlikely that the spectrum produced
by one source at $E<3\cdot 10^{15}eV$  would cut off at higher
energies, whereas another independent source at energies $E>3\cdot
10^{15}eV$ would cut off at lower energies, and they were joined
at the same energy $E\simeq 3\cdot 10^{15}eV$.

If the source is single, it would produce cosmic rays also in
other galaxies with characteristics similar to those observed in
the Galaxy. The aim of our work is to analyze the properties of
cosmic rays of superhigh energies $E>3\cdot 10^{18}eV$ observed on
the Earth, coming outside, from other galaxies, and compare them
with the properties of Galactic cosmic rays.

\section{Ultrahigh energy cosmic rays}

The energy distribution of the flux of cosmic rays $I(E)$ observed on the Earth at energies
$E > 3\cdot 10^{18}eV$ noticeably deviates from the power law distribution, which is
typical for lower energies, $3\cdot 10^{15}eV <E < 3\cdot 10^{18}eV$, where
$I(E)\propto E^{-3.1}$. This region is called 'ankle', here cosmic rays of
Galactic origin are replaced by particles, the origin of which has extragalactic
nature. They come from active galaxies located at large distances.
In particular, there is identification of coming particles with energies $E\simeq 10^{19}eV$ from galaxy Cen A,
located at distance $4.6 Mpc$ (Auger Collaboration, 2010), and correlation with active galactic nuclei (AGN) within $\simeq 100 Mpc$
(Auger Collaboration; 2007, 2008).
Charged particles (protons
or nuclei) of such energy, $\simeq 10^{19}eV$, have large cyclotron radius $r_L$ in intergalactic
magnetic field $B\simeq 10^{-9} G$,
$$
r_L=\frac{E}{ZeB}=11Z^{-1}\left(\frac{E}{10^{19}eV}\right)\left(\frac{B}{1nG}\right)^{-1} \, Mpc.
$$
The magnetic field changes its direction on the scale of the order of $1 Mpc$, i.e. correlation length of magnetic field equals
$l_c\simeq 1Mpc$. As a result a charged particle moving to our Galaxy, walking distance
$r$, deviates from the initial direction on the angle $\Delta\theta$,
$$
\Delta\theta=\left(\frac{l_c r}{r_L^2}\right)^{1/2}=0.09Z\left(\frac{E}{10^{19}eV}\right)^{-1}\left(\frac{B}{1nG}\right)
\left(\frac{l_c}{1Mpc}\right)^{1/2}\left(\frac{r}{1Mpc}\right)^{1/2}.
$$
Therefore, the propagation of cosmic rays of high energy from
sources, that are closer than $r< 100Mpc$, can be considered as
rectilinear in a good approximation, while from more distant
sources as diffusive. Let us find the distribution function
$N(t,{\bf r},E)$ of cosmic rays of high energy, $E\simeq
10^{19}eV$, over the space ${\bf r}$ and the energy $E$,
considering that charged particles come to the given point from
many sources in space located at points of ${\bf r}={\bf r}_i$
and having powers $Q_i(E)$. Considering the distribution of
sources in space as homogeneous in average, the particle
distribution function of $N$ can be considered as isotropic,
depending only on the distance $r$. In addition, due to the large
number of sources, the particle distribution also can be
considered stationary, $N=N(r,E)$. We discuss two cases: rectilinear
and diffusive propagation of particles.

\subsection{Rectilinear motion}

In this case, the equation for the distribution function of cosmic rays has the form
\begin{equation}
-c\frac{\partial N}{\partial r}+\frac{\partial}{\partial E}\left(\frac{dE}{dt}N\right)=\sum_i\frac{Q_i(E)}{4\pi r_i^2}\delta(r-r_i).
\end{equation}
Here $c$ is the velocity of the light, $\delta(x)$ is the Dirac
delta function. The value of $dE/dt$ is energy losses of
particles. They consist of two parts: losses connected with the
interaction of energetic particles with the relict radiation and
losses associated with the expansion of the Universe. To the
first, it is the so called GZK effect (Greizen, 1966; Zatsepin \& Kuz'min, 1966) - pion production in the
reaction $p +\gamma \rightarrow N + \pi$. The characteristic
particle energy of this process is $E_\pi=\mu c^2(1+\mu/m_p)m_p
c^2/2T\simeq 4\cdot 10^{20} eV$ ($m_p$ is the mass of the proton,
$\mu$ is the mass of the muon, $T$ the temperature of the relict
radiation). The interaction of a proton with a photon produces a
neutral $\pi$-meson, the interaction of a photon with a nucleus
can produce also charged $\pi$-mesons. At lower energies there
becomes significant the generation of electron-positron pairs, $p
+ \gamma = p + e^+ + e^-$. The characteristic particle energy of
this reaction is $E_e=m_e c^2 m_pc^2/T \simeq 2.1\cdot 10^{18} eV$
($m_e$ is the electron mass). Energy losses in photo-pion
reactions are well described by the expression (Stanev et al., 2000)
\begin{equation}
\left(\frac{d\epsilon}{dt^\prime}\right)_\pi=-(1+\epsilon)\exp(-\epsilon^{-1}).
\end{equation}
Here we introduced the dimensionless energy $\epsilon=E/E_\pi=E/4\cdot 10^{20}eV$ and the dimensionless time $t^\prime=ct/L$, where $L$ is
the characteristic distance, passing which a particle loses the energy of the order of its initial value, $L=13.7 Mpc$. 
Distances conveniently to measure in the same units, $r^\prime=r/L$.
The energy losses due to the generation of electron-positron pairs (see the paper by Berezinsky, Gazizov \& Grigorieva, 2006) can be
described approximately by the expression
\begin{equation}
\left(\frac{d\epsilon}{dt^\prime}\right)_e=-a\epsilon(\epsilon_e^{-2}+b\epsilon_e^{0.6})^{-1}\exp(-\epsilon_e^{-1}), \, \epsilon_e=E/E_e=190\epsilon.
\end{equation}
Here the constants $a$ and $b$ are $a=4.5\cdot 10^{-4}, \, b=8.4\cdot 10^{-3}$ correspondingly. Adiabatic particle losses associated with
the general expansion of the Universe are described by the Hubble law, $dE/dt = -H E$, where $H$ is the Hubble constant, $H=72 \, km s^{-1}/Mpc$.
In dimensionless variables adiabatic losses are
\begin{equation}
\left(\frac{d\epsilon}{dt^\prime}\right)_a=-\alpha\epsilon,  \, \alpha=3.3\cdot 10^{-3}.
\end{equation}
Here it should be noted that the Hubble expansion takes place for galaxies included to the local supercluster of galaxies,
of the order of $ 70 Mpc$ scale, gravitationally unbounded. However, the local group of galaxies, to which our Galaxy belongs, of
about $1.5 Mpc$ scale, already forms a gravitationally bound system, in which relative velocities differ considerably from the Hubble law.
Therefore, in the local group, the value of $\alpha$ can be significantly less than the above value (4).

The total losses of particle energy are equal to
\begin{equation}
\frac{d\epsilon}{dt^\prime}=-(1+\epsilon)\exp(-\epsilon^{-1})-a\epsilon(\epsilon_e^{-2}+b\epsilon_e^{0.6})^{-1}\exp(-\epsilon_e^{-1})-\alpha\epsilon.
\end{equation}
Although the value of $\alpha$ is small, the adiabatic energy losses become important for particles of not very high energies,
$\epsilon_e<(\alpha/a)^{1/2}\simeq 2.7, \, E<5.7\cdot 10^{18} eV$, when interaction with relict photons
becomes not significant.

It should also be mentioned that in addition to relict photons in the intergalactic environment, there exists also the light from galaxies,
and it contributes to the energy losses of cosmic rays. But, as shown by calculations, their contribution is small compared with that of the relict
radiation (Aloisio, Berezinsky \& Grigorieva, 2013).

Let us introduce the dimensionless values of the power of sources and the distribution function, $Q_i^\prime=Q_i L/c, \, N^\prime=N L^3$.
The dimensionless variables (with index $\prime$, which we omit further) will appear in the equation (1). We introduce also the variable
$\tau$ instead of $\epsilon$,
$$
\tau=\int_\epsilon^{\epsilon^\prime}\frac{dx}{\mid d\epsilon/dt\mid_{\epsilon=x}}.
$$
The value of $\tau$ is the time during which a particle loses its energy from the initial value $\epsilon^\prime$ to the current one
$\epsilon$. As a result, Equation (1) will take the form
\begin{equation}
\frac{\partial}{\partial r}\left(\mid\frac{d\epsilon}{dt}\mid N\right)-\frac{\partial}{\partial\tau}\left(\mid\frac{d\epsilon}{dt}\mid N\right)=
-\sum_i\frac{Q_i(\epsilon)|\frac{d\epsilon}{dt}|}{4\pi r_i^2}\delta(r-r_i).
\end{equation}
Integrating equation (6) over $r$ from $0$ to $r_m$ with the constant sum $r+\tau=const$, we get
\begin{equation}
N(r=0,\epsilon)=\frac{1}{|d\epsilon/dt|}\sum_i\frac{Q_i(\epsilon_i)\mid\frac{d\epsilon}{dt}\mid_{\epsilon=\epsilon_i}}{4\pi r_i^2}+
\frac{\mid d\epsilon/dt\mid_{\epsilon=\epsilon_m}}{\mid d\epsilon/dt\mid}N(\epsilon_m, r_m).
\end{equation}
The value of $r_m$ is the distance to which the approximation of rectilinear propagation is valid, $r_m\simeq 100Mpc/13.7Mpc=7.3$. The energy
$\epsilon_m$ is determined by the ratio
$$
r_m=\int_\epsilon^{\epsilon_m}\frac{dx}{\mid d\epsilon/dt\mid_{\epsilon=x}}.
$$
In Equation (7) the first term is the contribution of the sources of cosmic rays of superhigh energies, located at distances $r<r_m$
from the Galaxy, whose particles are observed in the Galaxy ($r=0$). Because, according to analysis by Takami \& Sato (2009), the density of such sources is
of $10^{-2}-10^{-4} Mpc^{-3}$, and the number of these sources in the sphere of radius $r_m$ is significantly greater than unity, it is possible
to go from the summation to the
integration over volume, $4\pi\int_0^{r_m} r_i^2 dr_i$, introducing the average density of sources ${\bar q}(\epsilon)$. The value of
energy, on which the quantity ${\bar q}(\epsilon)$ depends, is the initial energy of a particle, which overcomes the distance $r_i$ from the source to
the Galaxy, $\epsilon=\epsilon_i$,
$$
r_i=\int_\epsilon^{\epsilon_i}\frac{dx}{\mid d\epsilon/dt\mid_{\epsilon=x}}.
$$
However, at constant energy $\epsilon$ the distance $r_i$ and the energy $\epsilon_i$ are connected, so the integration over $r_i$ can be
replaced by the integration over $\epsilon_i$. The result is
$$
\frac{1}{|d\epsilon/dt|}\sum_i\frac{Q_i(\epsilon_i)\mid\frac{d\epsilon}{dt}\mid_{\epsilon=\epsilon_i}}{4\pi r_i^2}=
\frac{\int_\epsilon^{\epsilon_m} {\bar q}(\epsilon^\prime)d\epsilon^\prime}{|d\epsilon/dt|}.
$$
Thus, the distribution of cosmic rays of high energy, $E>3\cdot 10^{18}eV$, observed in the Galaxy, is largely determined by
the loss function  $|d\epsilon/dt|$ (5) in the intergalactic space, but not by the sources, distribution of
which over energy ${\bar q}(\epsilon)$ is in average form.

\subsection{Diffusive motion}

In order to determine the contribution of distant ($r>r_m$) sources into the distribution of cosmic rays observed in our Galaxy
(the second term in Equation (7)) we have to consider the diffusion region $r>r_m$. The equation for the distribution function $N(r,\epsilon)$
is
\begin{equation}
\frac{\partial}{\partial E}\left(\frac{dE}{dt}N\right)-\frac{D}{r^2}\frac{\partial}{\partial r}\left(r^2\frac{\partial N}{\partial r}\right)
=\sum_i\frac{Q_i(E)}{4\pi r_i^2}\delta(r-r_i).
\end{equation}
Here $D$ is the coefficient of diffusion of cosmic rays in the intergalactic space. It, as well as in the Galaxy, depends on the particle energy.
Let us suppose that this dependence is a power law, $D=D_0 E^\kappa$. For cosmic rays in the Galaxy $\kappa=0.7$.
Also as before, we introduce the dimensionless variables: the distances are
measured in $L$, time - in $L/c$, the energy - in $4\cdot 10^{20}eV$, the diffusion coefficient - in units of
$D_0^\prime=D_0(4\cdot 10^{20}eV)^{\kappa}/Lc$.
We introduce also the effective time $\tau_D$,
$$
\tau_D=\int_\epsilon^{\epsilon^\prime}\frac{x^{\kappa}dx}{\mid d\epsilon/dt\mid_{\epsilon=x}}.
$$
As a result, Equation (8) becomes
\begin{equation}
\frac{\partial}{\partial\tau_D}\left(\mid\frac{d\epsilon}{dt}\mid N\right)-\frac{D_0}{r^2}\frac{\partial}{\partial r}
\left(r^2\frac{\partial}{\partial r}\left(\mid\frac{d\epsilon}{dt}\mid N\right)\right)=
\epsilon^{-\kappa}\mid\frac{d\epsilon}{dt}\mid\sum_i\frac{Q_i(\epsilon)}{4\pi r_i^2}\delta(r-r_i).
\end{equation}
Knowing the Green function of the equation of diffusion in a spherical region $0<r<\infty$ (the diffusion approximation includes also the
region $0<r<r_m$ of the scale of the free path length),
\begin{equation}
G(r,r^\prime,t)=\frac{1}{(4\pi D_0t)^{1/2}}\frac{r^\prime}{r}\left\{\exp[-\frac{(r-r^\prime)^2}{4D_0t}]-\exp[-\frac{(r+r^\prime)^2}{4D_0t}]\right\},
\end{equation}
integrating it with the right hand part of Equation (9) over $r^\prime$ and $\tau_D$, then, as before, going from summation over $i$ to
the integration over $r_i$ from $r_m$ to infinity,
introducing again the density of sources ${\bar q}(\epsilon)$, and transforming the integration over $\tau_D$ to the integration over
$\epsilon^\prime$, we get
\begin{equation}
N(r_m,\epsilon)=\frac{1}{|d\epsilon/dt|}\int_\epsilon^{\epsilon_0}{\bar q}(\epsilon^\prime)
\Phi\left[\left(\frac{r_m^2}{4D_0\tau_D(\epsilon^\prime)}\right)^{1/2}\right]d\epsilon^\prime.
\end{equation}
Here the function $\Phi(y)$ is equal to
$$
\Phi(y)=2-\frac{2}{\pi^{1/2}}\int_0^{2y}\exp(-z^2)dz+\frac{1}{\pi^{1/2}y}[1-\exp(-4y^2)].
$$
It varies monotonically from the value $\Phi=2$ at $y=0$ to $\Phi=1$ for $y>>1$. And the fast transition of $\Phi$ to the value $\Phi\simeq 1$ occurs
already at $y\simeq 2$ (see Figure 1). The typical value of $y^2$ is the ratio of the length of the mean free path length of particles $r_m$ to the path
$L$, travelling along which a
particle loses a significant part of its energy, $y^2\simeq r_m/L = 7.3$. In addition, large time
$\tau_D$, small $y$, corresponds to high initial energies of particle $\epsilon^\prime$. Because the
source power ${\bar q}(\epsilon)$ decreases with energy, and, as we will see, rather fast, ${\bar q}(\epsilon)\propto\epsilon^{-2.7}$, then
the contribution of the function $\Phi(y)$ at small values of its argument into the integral (11) is small.
Thus, we can consider $\Phi\simeq 1$ in the expression (11).
Altogether, the required distribution function of particles on the boundary of $r=r_m$ at $\epsilon=\epsilon_m$ equals
\begin{equation}
N(r_m,\epsilon_m)=\frac{1}{\mid d\epsilon/dt\mid_{\epsilon=\epsilon_m}}\int_{\epsilon_m}^{\epsilon_0}{\bar q}(\epsilon^\prime)d\epsilon^\prime.
\end{equation}
The value of $\epsilon_0$ is the maximum energy of particles in sources.

\begin{figure}
\begin{center}
\includegraphics[width=8cm]{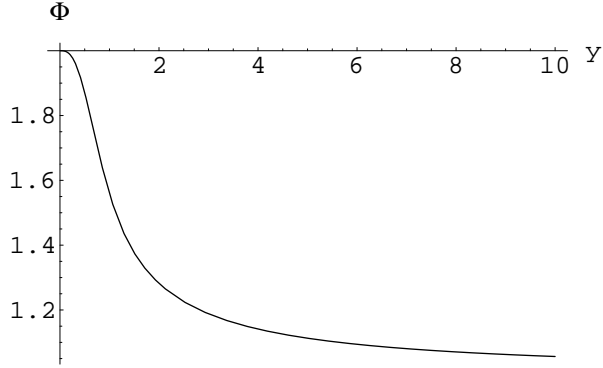}
\end{center}
\caption {The function $\Phi(y)$ of the parameter $y=(r_m^2/4D_0\tau_D)^{1/2}$. See formula (11).}
\end{figure}

\section{Energy distribution of extragalactic particle observed in the Galaxy}

Substituting the resulting distribution (12) into the expression (7), we obtain
the sought distribution function of extragalactic cosmic rays,
observed in the Galaxy,
\begin{equation}
N(r=0,\epsilon)=\frac{\int_\epsilon^{\epsilon_0}{\bar q}(\epsilon^\prime)d\epsilon^\prime}{|d\epsilon/dt|}.
\end{equation}
Equation (13) has a simple physical meaning: the particle flux in the energy
space, $N|d\epsilon/dt|$, is equal to the total flux produced by
sources $Q_i$. Moreover, since we have many sources, their fluxes are summing,
ranging from close sources, providing a full range of energies
from $\epsilon=\epsilon_{min}$ up to the maximum one $\epsilon=\epsilon_0$,
to distant sources from which we observe only large initial energy
$\epsilon\simeq\epsilon_0$. Suppose that the distribution ${\bar q}(\epsilon)$
is the power law function, ${\bar q}(\epsilon)=q_0 \epsilon^{-\beta}$.
Then the distribution function $N(\epsilon)$ is equal to
\begin{equation}
N(\epsilon)=\frac{q_0}{\beta-1}\frac{\epsilon^{-\beta+1}-\epsilon_0^{-\beta+1}}{(1+\epsilon)\exp(-1/\epsilon)+
a\epsilon(\epsilon_e^{-2}+b\epsilon_e^{0.6})^{-1}\exp(-1/\epsilon_e)+\alpha\epsilon}.
\end{equation}
The graph of the function $\epsilon^3 N(\epsilon)$ for
$\epsilon_0=25, \, E_0=10^{22}eV$, and $\beta=2.7$ is shown on Figure 2. One can see
that at energies $E< 10^{18} eV$ and $E>4\cdot 10^{20} eV$ the
distribution $N(\epsilon)$ is the power law, $N(\epsilon)\propto
\epsilon^{-\beta}$. The same slope is observed at the intermediate
energies, in the region of the maximum relative losses of energy,
$(dE/dt)/E$, on the birth of electron-positron pairs, $E\simeq
10^{19} eV$. In the region $0.2<\epsilon<1$, $N$ grows
exponentially with decreasing of energy,
$N(\epsilon)=q_0\epsilon^{-\beta+1}\exp(1/\epsilon)/(\beta-1)$.
According to observations summarized by Berezinsky (2013), the
flux of particles $I(\epsilon)=cN(\epsilon)/4\pi$ in the energy
range $6\cdot 10^{18}eV<E<4\cdot 10^{19}eV,\,1.5\cdot
10^{-2}<\epsilon<0.1$, is indeed the power law function of energy
with the index $\beta=2.7$. When analyzing the data of
observations it is convenient to use the function $\epsilon^3
N(\epsilon)$, which clearly describes the transition from
Galactic distribution ($E^3 I(E)\propto E^{-0.1}$) for $E<3\cdot
10^{18}eV$ to the extragalactic one ($E^3 I(E)\propto E^{0.3}$)
for $6\cdot 10^{18}eV<E<4\cdot 10^{19}eV$. The function
$\epsilon^3 N(\epsilon)$ has a maximum at
$\epsilon=\epsilon_1=0.1,\, E_1=4\cdot 10^{19}eV$ (see Figure 2).
At energies $\epsilon>\epsilon_1$ the function $\epsilon^3
N(\epsilon)$ exponentially decreases not to zero, but to the
minimum value at $\epsilon_2$, $\epsilon_2\simeq 1,\, E_2\simeq
4\cdot 10^{20}eV$. After that the distribution $N(\epsilon)$
continues to fall down by a power law manner with the same index
$\beta$, $N(\epsilon)\propto\epsilon^{-\beta}$. The fall down of
the function $\epsilon^3 N(\epsilon)$ at energy $\epsilon_1$ to
its value at $\epsilon_2$ is approximately one and half orders.
The strong growth of the density of particles for
$\epsilon<\epsilon_2$ is explained by the sharp decreasing of the
rate of energy losses of particles when braking by relict photons
becomes small. Here particles are accumulated. The distribution of
$N(E)$ for energies $E>4\cdot 10^{20}eV$ does not go exponentially
to small values that can seem as the result of the GZK effect.
Here we observe that the distribution of $N(E)$ repeats the
distribution of extragalactic sources ${\bar q}(E)$, unless of the
maximum energy $E_0$ is not close to the energy $4\cdot
10^{20}eV$. However, at energies $E>4\cdot 10^{20}eV$ the
observational data have great uncertainty, which does not allow to
make a conclusion about the growth of the function $E^3 N(E)$ at
high energies $E>4\cdot 10^{20}eV$ under the condition $E_0>4\cdot
10^{20}eV$. The energy range $E<3\cdot 10^{18}eV$ (see Figure 2),
where the distribution $N(E)$ also reproduces the distribution of
extragalactic sources ${\bar q}(E)$, is hidden by Galactic cosmic
rays.

\begin{figure}
\begin{center}
\includegraphics[width=8cm]{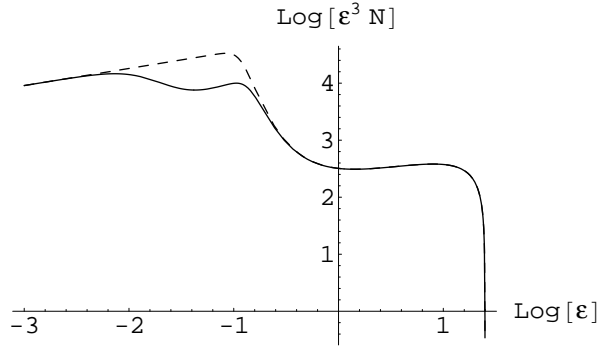}
\end{center}
\caption {The function $\epsilon^3 N(\epsilon)$, $N(\epsilon)$ is given by the formula (14). Here we choose the value
of maximum energy$ E_0=10^{22} eV, \, \epsilon_0=25$, and the index $\beta=2.7$.
The dashed line represents this function when the electron-positron pairs production is absent.}
\end{figure}

Estimate now the average power density of sources ${\bar q}$.
Observations give the value of the extragalactic particle flux, where
there is maximum
value of product $E^3 I(E)$ at the energy $E_1=4\cdot 10^{19}eV$,
$$
I(E_1)=4.7\cdot 10^{-39} cm^{-2}s^{-1}sr^{-1}eV^{-1}.
$$
Accordingly, the density of particles $N(E)$, $N(E)=4\pi I(E)/c$, is
$$
N(E_1)= 5.9\cdot 10^{25}Mpc^{-3}eV^{-1}.
$$
On the other hand, the dimensionless density of particles $N(\epsilon_1)$ and
average power of sources ${\bar q}(\epsilon_1)$ are connected by
the relation (14),
$$
{\bar q}(\epsilon_1)=(\beta -1)\left[(\epsilon_1^{-1}+1)\exp(-1/\epsilon_1)+a(\epsilon_{e1}^{-2}+b\epsilon_{e1}^{0.6})^{-1}
\exp(-1/\epsilon_{e1})+\alpha\right] N(\epsilon_1)=2\cdot 10^{-2}N(\epsilon_1).
$$
Bearing in mind that the dimensional quantity ${\bar q}$ is $c/L=7.1\cdot 10^{-16}s^{-1}$ times less than the dimensionless one, we get
$$
{\bar q}(E_1)=8.4\cdot 10^{8}eV^{-1}s^{-1}Mpc^{-3}.
$$
Considering that, as well as in our Galaxy, the power law spectrum with index $\beta=2.7$ continues up to small energies $E_{min}\simeq 5GeV$, we get
\begin{equation}
{\bar q}(E)= 8.4\cdot 10^{8}\left(\frac{E}{4\cdot 10^{19}eV}\right)^{-2.7}eV^{-1}s^{-1}Mpc^{-3}.
\end{equation}
The total average power density of extragalactic sources ${\cal E}=\int_{E_{min}}^{E_0}{\bar q}(E)EdE$ is equal to
\begin{equation}
{\cal E}=1.8\cdot 10^{43}\left(\frac{E_{min}}{5GeV}\right)^{-0.7}erg \, s^{-1}Mpc^{-3}.
\end{equation}
This value divided by the density of galaxies, $N_g\simeq 1Mpc^{-3}$, at least two orders of magnitude exceeds the power of the Galaxy in cosmic rays,
$10^{41}erg \, s^{-1}$. If we assume that the sources of ultrahigh energy particles are active galactic nuclei with relativistic
jets generated inside, the power of which is $(10^{45} - 10^{46}) erg \, s^{-1}$ (Mao-Li, et al., 2008), then their density in the Universe is of
$N_g\simeq(10^{-2} - 10^{-3}) Mpc^{-3}$. The same estimation of the density of extragalactic sources of cosmic rays follows also from the
conditions of isotropy of arrival of particles in the range of energies $E\simeq 10^{19}eV$ (Abbasi et al., 2004).

\section{Galactic cosmic rays}

The most surprising fact, following from the previous
consideration, is that the spectrum of the power density of extragalactic cosmic rays
(15) has the the same slope that have Galactic cosmic rays density before
the 'knee', $E<3\cdot 10^{15}eV$, $\beta=2.7$. Let us note here that the cosmic rays
power density ${\bar q}(E)$ understanding as the average power radiated by a galaxy as a 
whole is proportional to the density of cosmic rays in a galaxy $N(E)$, ${\bar q}(E)\propto
N(E)Sc$, where the value of $S$ is the galactic surface radiated cosmic rays.
This indicates that
the nature of the origin of cosmic rays in the Galaxy and in
active galactic nuclei is the same. The formation of the spectrum
with index $\beta=2.7$ in the Galaxy is explained by the fact that
the index of the power law energy spectrum of source is of
universal value of $\beta=2$. This is valid as for the acceleration
of particles on fronts of strong shock waves (Krymskii, 1977; Bell, 1978), as for the
acceleration at the base of the jets emitted near massive black
holes in centers of galaxies (Istomin, 2014). Next, accelerated particles,
spreading over a galaxy, are scattered by inhomogeneities of a
magnetic field. Their motion becomes diffusive. Moreover, the
coefficient of diffusion is larger for particles with larger
energies, $D\propto E^{0.7}$ (Ptuskin, 2007). The density of particles is
equal to the product of the power of a source $Q$ to the lifetime
of particles $\tau$, $N=Q\tau$. The lifetime is the time of escape
of particles from a galaxy, $\tau=R^2/D$, $R$ is the radius of a
galaxy. Thus, $N(E)\propto E^{-2-0 .7}$. There arises the
question: why in our Galaxy the spectral index of cosmic rays at
energies $E>3\cdot 10^{15}eV$ deviates from the value of
$\beta=2.7$? Spectrum becomes softer, $\beta =3.1$. If the source
of cosmic rays in a galaxy is strong shock waves from supernova
explosions, it is not clear why the spectrum of high energy
particles from active galaxies remains with index $\beta=2.7$ up
to energies $\sim 10^{20}eV$. More natural to assume that the
source of cosmic rays is a jet emitted from active galactic nucleus
and whose power significantly exceeds the power transformed to
particles by supernova explosions. Istomin (2014) suggested that
Galactic cosmic rays were produced by the jet, emitted from the
center of the Galaxy. Giant bubbles of relativistic gas, observed
above and below the Galactic plane, are remnants of this bipolar
jet existed previously. From the size of the bubbles it follows
that the jet switched on $2.4\cdot 10^7$ years ago, and it worked
at least $10^7$ years. Thus, before the jet switched off the
Galaxy and its halo were uniformly filled by cosmic rays with
the spectral index $\beta=2.7$ (Istomin, 2014). After the source turned off
particles continue to flow out from the Galaxy, and their density
begins to decrease with time. Consider how it is happen.

Suppose that at time $t=0$ cosmic rays with density $N_0(E)$ uniformly filled
the spherical region (Galaxy and halo) of radius $R$. Then, assuming that the
motion of particles is of diffusion
character, using the Green function (10), we find the density of cosmic
rays $N(r,E,t)$ at the point located
at distance $r$ from the center of the Galaxy and at the time $t$,
$N(r,E,t)=N_0(E)F(p,r)$,
\begin{eqnarray}
&&F(p,r)=\frac{1}{\pi^{1/2}}\left\{\frac{R}{2rp}\left[\exp[-p^2(1-r/R)^2]-\exp[-p^2(1+r/R)^2]\right] \right. \nonumber  \\
&&\left. +\int_{-p(1-r/R)}^{p(1+r/R)}\exp(-y^2)dy\right\},
\end{eqnarray}
where the parameter $p$ is equal to $p(E,t)=[R^2/4D_g(E)t]^{1/2}$.
For small values of $p$, $p<<1$, the function $F$ is equal to
$F=4p/\pi^{1/2}$, for large values $p>>1$, $F=1$. The graph of the function
$F(p)$ for two different values of $r=R/2$
and $r=0$ are presented on the Figure 3.
\begin{figure}
\begin{center}
\includegraphics[width=8cm]{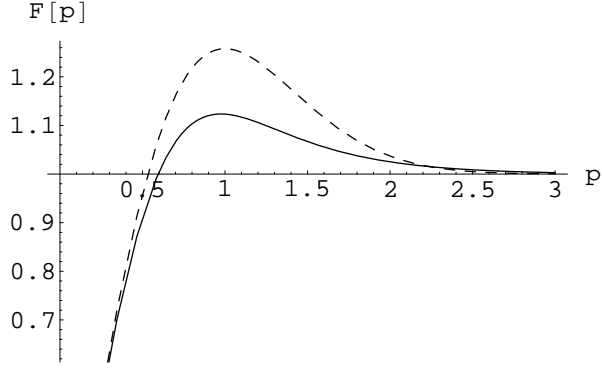}
\end{center}
\caption {The function $F(p)$. The solid line corresponds to $r=R/2$, the dashed line - $r=0$}
\end{figure}
It is seen that the transition of
density of cosmic rays from the initial
distribution $N_0$ ($p>1$) to the falling down with time,
$N=N_0(4R^2/\pi D_gt)^{1/2}$, takes place
at $p\simeq 1$. At this point there is a transition of the spectrum of cosmic
rays from the original,
$N\propto E^{-2.7}$, with energies $E<E_k$ to the distribution of
$N\propto E^{-2.7-0.7/2}=E^{-3.05}$ for $E>E_k$.
The value of energy $E_k$ is defined by the equality $p=1$, $D_g(E_k)=R^2/4t$.
In the center of the Galaxy ($r=0$) the transition
occurs exactly at the point $p=1$, for us ($r\simeq R/2$), this point is also
very close to unity.
When $p=1$ there is a local increase of density of cosmic rays, i.e. near
energies $E\simeq E_k$ the
distribution $N(E)$ is really observed as 'knee' (see Fig. 4). The position
of the 'knee'
depends on the time $t$, passed after turning off of the source. Knowing the
position of the 'knee' at the present time $t=t_0$,
we find
$$
t_0=\frac{R^2}{4D_g(E_k)}=4.2\cdot 10^4\left(\frac{R}{5\cdot 10^{22}cm}\right)^2\left(\frac{D_g(1GeV)}{2.2\cdot 10^{28}
cm^2s^{-1}}\right)^{-1}\left(\frac{E_k}{3\cdot 10^{15}eV}\right)^{-0.7}year.
$$
It should be noted that 'knee' is moving with time,
$E_k=3\cdot 10^{15}(t/t_0)^{-1.43}eV$.
The speed is now equal to $dE_k/dt=-1.43 E_k/t_0=-10^{11}eV/year$.
\begin{figure}
\begin{center}
\includegraphics[width=8cm]{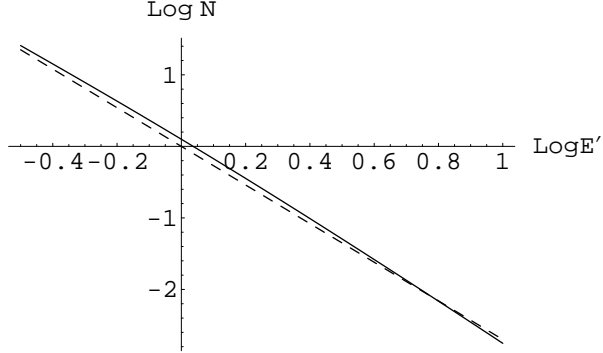}
\end{center}
\caption {The distribution of Galactic cosmic rays $N(E)\propto E^{-2.7}F(p)$ over
energy $E$ near the 'knee' at the point $r=R/2$, $E'=E/E_k$. The function $F(p)$ is defined by the expression (17). The
dashed line is the dependence $N(E)\propto E^{-2.7}$.}
\end{figure}
\section{Conclusions}
We showed that the distribution of power of sources of
extragalactic cosmic rays in the energy range of $3\cdot
10^{18}eV<E<10^{21}eV$ is a power law, ${\bar q}(E)\propto
E^{-2.7}$. Thus, it is the same as in our Galaxy at the energy
range below the 'knee', $E<3\cdot 10^{15}eV$. This indicates the
common nature of the origin of cosmic rays in the Galaxy and in
other galaxies. However, the power of extragalactic sources at
least two orders of magnitude exceeds the capacity of the Galaxy
(see formula (16)). The conclusion from this is that a 'normal'
galaxy, to which belongs our Galaxy, is not the source of cosmic
rays. The estimation of the density of extragalactic cosmic rays
sources, $N_g\simeq 10^{-2}-10^{-3} \, Mpc^{-3}$, indicates
galaxies with active nuclei. The possible source of energetic
charged particles is relativistic jets emitted from surroundings
of massive black holes. Thus, acceleration by shock waves from
supernova explosion is not possible to explain the origin of
cosmic rays up to energies $\simeq 10^{18}eV$. Otherwise, all
galaxies in more or lesser degree would are sources of cosmic
rays, because in all of them there are explosions of supernova.
Why, nevertheless, we observe cosmic rays in the Galaxy is
explained that once in the past the Galaxy was also active. From
the center of the Galaxy there emitted the relativistic bipolar
jet, the remnants of which is observed now above and below the
Galactic plane as bubbles of relativistic gas (Su, Slatyer \&
Finkbeiner, 2010). Due to the fact that now this source of cosmic
rays in the Galaxy is not working, it appears the deflection of
the spectrum of Galactic cosmic rays from dependence $N(E)\propto
E^{-2.7}$ in the whole energy range. For $E>3\cdot 10^{15}eV$, the
spectrum becomes softer, $N(E)\propto E^{-3.1}$. This is because
particles, that once filled the Galaxy, leave it the faster the
larger their energy, because their diffusion coefficient increases
with energy, $D\propto E^{0.7}$. Particles of larger energies
quickly leave the Galaxy than particles less energies. For the
diffusive motion of particles $N(E)\propto
N_0(E)(Dt)^{-1/2}\propto E^{-2.7-0.7/2}$ for energies $E>E_k$ and
$N(E)\simeq N_0(E)$ for energies $E<E_k$, where $N_0(E)$ is the
initial distribution of particles. Thus, the 'knee' formation
reflects the escape of particles from the Galaxy. Knowing the
position of the 'knee' now one can estimate the time when the
source of cosmic rays in the Galaxy stopped, it occurred
$4.2\cdot 10^{4}$ years ago. The position of the 'knee' is not
constant in time, it must move, $E_k\propto t^{-1.43}$. This
motion one can notice if to have a sufficient accuracy of the
measurement of the 'knee' position. During 50 years the change in
the position of the 'knee' is $\Delta E_k=-5\cdot 10^{12}eV$.

\section*{Aknowlegements}
Author thanks V. Berezinsky for the fruitful discussions.

This work was done under support of the Russian Foundation for Fundamental
Research (grant numbers 14-02-00831 and 13-02-12103).

\end{document}